\documentclass[
    reprint,
    superscriptaddress,
    amsmath,
    amssymb,
    aps,
    nofootinbib
]{revtex4-2}

\usepackage[T1]{fontenc}
\usepackage[utf8]{inputenc}

\usepackage{microtype}

\usepackage{amsmath}
\usepackage{amssymb}
\usepackage{amsthm}
\usepackage{mathtools}

\usepackage{graphicx}
\usepackage{booktabs}
\usepackage{multirow}

\usepackage{subfig}

\usepackage{enumitem}

\usepackage{xcolor}

\usepackage{tikz}
\usetikzlibrary{
    positioning,
    arrows.meta,
    shapes.geometric,
    calc
}

\usepackage[
    colorlinks=true,
    linkcolor=blue,
    citecolor=blue,
    urlcolor=blue
]{hyperref}

\graphicspath{{output/}}

\begin{document}

\title{
High-Dimensional Concentration and Retrieval Instability in Embedding Spaces:\\
Implications for Retrieval-Augmented Generation}

\author{Ernesto Lopez Fune}
\email{ernestolopezfune@gmail.com}
\affiliation{Data Empirix, Paris, France}

\date{\today}

\begin{abstract}
Embedding-based retrieval systems rely on the assumption that geometric proximity in high-dimensional representation spaces reflects semantic relevance. However, high-dimensional geometry induces concentration phenomena that can reduce the discriminative power of similarity measures and can destabilize nearest-neighbor retrieval. This work studies distance concentration, cosine concentration, contrast collapse, hubness, and retrieval instability through controlled numerical experiments across multiple synthetic distributions. The results show that similarity signals progressively lose contrast as dimension increases, leading to unstable retrieval behavior and structural bias in nearest-neighbor selection. A simplified Retrieval-Augmented Generation experiment further suggests that these effects can degrade grounding reliability upstream of generation. These findings motivate geometry-aware diagnostics and robustness-oriented retrieval strategies for embedding-based AI systems. The experiments are intentionally synthetic in order to isolate intrinsic geometric effects.

\end{abstract}

\maketitle



\section{Introduction}

Large Language Models (LLMs) and Retrieval-Augmented Generation (RAG) systems rely heavily on high-dimensional vector representations. Textual data is mapped into embedding spaces where semantic similarity is approximated through geometric proximity. This paradigm underlies modern semantic search engines, vector databases, recommender systems, and retrieval-based generative pipelines.

A key operational assumption in these systems is that nearby vectors correspond to semantically relevant information. In practice, retrieval mechanisms rely on nearest-neighbor search using similarity measures such as cosine similarity or Euclidean distance.

However, high-dimensional spaces exhibit several well-known geometric phenomena. Classical results show that pairwise distances tend to concentrate, relative contrast decreases, and certain points become disproportionately frequent nearest neighbors, a phenomenon known as hubness \cite{beyer1999nearest, aggarwal2001surprising, radovanovic2010hubs}. These effects can reduce the discriminative power of similarity-based retrieval. More broadly, these effects are related to the classical ``curse of dimensionality'' originally identified by Bellman in the context of dynamic programming and high-dimensional optimization~\cite{bellman1961adaptive}.

In parallel, modern RAG systems have introduced renewed interest in retrieval reliability and grounding quality \cite{lewis2020rag, karpukhin2020dpr, gao2023rag}. Since generative models operate on retrieved context rather than on the entire knowledge base, instability or degradation in the retrieval layer can propagate downstream and affect generated outputs.

This work studies how high-dimensional concentration phenomena affect retrieval behavior under controlled experimental conditions. Rather than focusing on a particular embedding model, we isolate intrinsic geometric effects using synthetic data drawn from multiple statistical distributions. This approach allows us to separate concentration effects from model-specific artifacts such as corpus bias, anisotropy, or training dynamics.

We analyze several complementary quantities, such as: pairwise cosine similarity, variance of cosine similarity, coefficient of variation of distances, relative contrast, cosine score gaps, nearest-neighbor instability, and hubness metrics. We further introduce a simplified retrieval experiment inspired by RAG pipelines to study how concentration-induced instability can affect retrieval success and weak grounding proxies.

This paper makes the following contributions. First, it provides a unified empirical study of distance concentration, cosine concentration, relative contrast collapse, hubness, and retrieval instability in high-dimensional embedding-like spaces. Second, it compares raw and standardized representations across multiple distributional regimes, allowing distributional bias and intrinsic concentration effects to be separated. Third, it introduces retrieval-oriented diagnostics, including cosine score gaps, perturbation stability, and hubness measures, as operational indicators of nearest-neighbor reliability. Finally, it presents a simplified RAG-inspired experiment illustrating how high-dimensional concentration can weaken retrieval success upstream of generation.


\begin{figure*}[htbp]
\centering
\begin{tikzpicture}[
    node distance=1.6cm,
    block/.style={
        rectangle,
        rounded corners,
        draw=black!55,
        fill=gray!8,
        align=center,
        minimum width=3.8cm,
        minimum height=0.9cm,
        font=\small
    },
    arrow/.style={
        -{Latex[length=2.5mm]},
        thick
    }
]

\node[block] (hd) {High-dimensional\\embedding space};

\node[block, right=of hd] (conc) {
Distance and cosine\\concentration
};

\node[block, right=of conc] (gap) {
Score-gap collapse\\and hubness
};

\node[block, below=of gap] (instab) {
Retrieval instability\\under perturbations
};

\node[block, left=of instab] (weak) {
Weak or incomplete\\retrieved context
};

\node[block, left=of weak] (rag) {
Potential degradation\\of grounding
};

\draw[arrow] (hd) -- (conc);
\draw[arrow] (conc) -- (gap);
\draw[arrow] (gap) -- (instab);
\draw[arrow] (instab) -- (weak);
\draw[arrow] (weak) -- (rag);

\end{tikzpicture}

\caption{
Conceptual pathway from high-dimensional concentration to retrieval instability and potential degradation of grounding. The diagram does not imply that geometry is the sole cause of hallucination, but identifies one structural mechanism that can weaken retrieval quality upstream of generation.
}

\label{fig:conceptual_pathway}

\end{figure*}
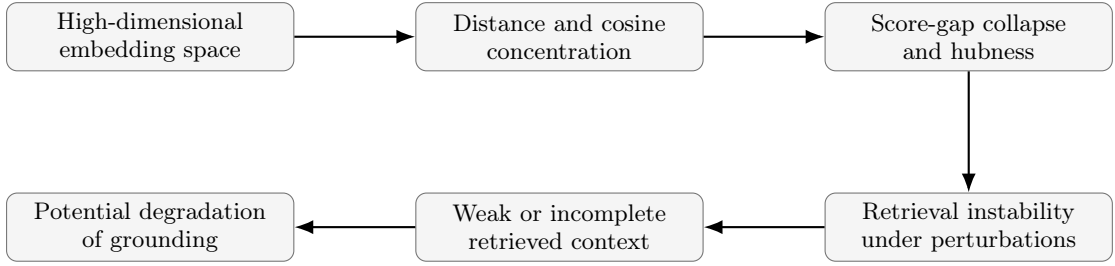

\section{From Language Models to High-Dimensional Geometry}\label{sec:foundations}
Modern AI systems rely on a fundamental abstraction: textual data is represented as vectors in high-dimensional continuous spaces. This transformation enables the application of numerical computation to inherently discrete and symbolic information, forming the basis of embedding-based methods used across contemporary machine learning pipelines.

In the context of LLMs, these vector representations play a central role in both retrieval and generation processes. In particular, the quality of retrieved information, and consequently the reliability of generated outputs, depends critically on the geometric structure of the underlying embedding space. Understanding how similarity, distance, and neighborhood relationships behave in high dimension is therefore essential for analyzing the performance and limitations of such systems.

This section adopts a constructive perspective. We first introduce the representation of text through embeddings in LLM-based architectures. We then present the geometric tools used to compare vectors, including scalar products, norms, distances, and angular similarity. Finally, we formalize the statistical phenomena that arise in high-dimensional regimes, which will serve as the theoretical foundation for the experimental analysis presented in the following sections.

\subsection{Text Representation in High Dimensions}\label{subsec:embeddings}
In modern LLM-based systems, textual inputs are transformed into dense numerical representations, commonly referred to as \emph{embeddings}. These representations are constructed so that semantic relationships between texts are reflected through geometric proximity in the embedding space.

In practice, embedding dimensions typically range from a few hundred to several thousand (e.g., $m \in \{384, 768, 1536, 3072\}$), depending on the underlying model architecture and design trade-offs between expressivity and computational efficiency. For example, widely used transformer-based embedding models operate in dimensions that reflect both architectural design and computational trade-offs. Early encoder models such as BERT produce embeddings of dimension $m=768$ \cite{vaswani2017attention}, which has become a de facto baseline in many applications. More recent systems employ higher-dimensional representations to capture richer semantic structure: OpenAI embedding models typically use $m=1536$, while Google Gemini embeddings can reach up to $m=3072$, with configurable lower-dimensional outputs \cite{openai_embeddings, google_gemini_embeddings}. Overall, modern embedding pipelines commonly operate in the range $m \in [768, 3072]$, balancing expressivity with efficiency. As will be shown in the following sections, operating in such high-dimensional regimes has non-trivial geometric and statistical consequences that directly impact similarity-based retrieval. 

To illustrate, consider the following example using a standard embedding model such as those provided by OpenAI or similar systems. The sentences:
\begin{quote}
\emph{``The cat is sleeping on the sofa.''} \\
\emph{``A dog is resting on the couch.''}
\end{quote}
are mapped to vectors that exhibit a high cosine similarity, reflecting their semantic proximity. In contrast, a sentence such as:
\begin{quote}
\emph{``The stock market experienced a significant downturn.''}
\end{quote}
is mapped to a representation that lies significantly farther away in the embedding space.

This behavior underpins a wide range of downstream tasks, including semantic search, document retrieval, clustering, and recommendation systems. In all these settings, the operational assumption is that geometric proximity provides a reliable proxy for semantic relevance.

However, as will be discussed in the following subsections, this assumption becomes increasingly fragile in high-dimensional regimes, where geometric and statistical effects may distort similarity relationships.

\subsubsection{Chunking, Tokenization and Similarity-Based Retrieval}
In RAG pipelines, raw documents are first decomposed into smaller, manageable units. This process begins with \emph{tokenization}, where text is split into elementary units (tokens) that can be processed by the model. These tokens are then grouped into contiguous segments, referred to as \emph{chunks}, typically comprising between $100$ and $1000$ tokens depending on the application and context window constraints.

Each chunk is subsequently mapped to an embedding independently:
\begin{equation}
\text{chunk}_i \longrightarrow \vec{x}_i \in \mathbb{R}^m.
\end{equation}

As a result, a document collection
\begin{equation}
\mathcal{C} = \{\text{chunk}_1, \text{chunk}_2, \ldots, \text{chunk}_N\}
\end{equation}
is represented as a finite set of vectors
\begin{equation}
\mathcal{X} = \{\vec{x}_1, \vec{x}_2, \ldots, \vec{x}_N\} \subset \mathbb{R}^m,
\end{equation}
which can be interpreted as a discrete point cloud in a $m-$dimensional space.

Given a query $\text{chunk}$, its embedding $\vec{x} \in \mathbb{R}^m$ is compared to the collection $\mathcal{X}$, and the most relevant chunks are retrieved via a similarity-based ranking:
\begin{equation}
\mathrm{Top}_k(\vec{x}) = \arg\max_{\vec{x}_i \in \mathcal{X}} \ \mathrm{sim}(\vec{x}, \vec{x}_i).
\end{equation}

This nearest-neighbor retrieval formulation is closely related to classical information retrieval systems, where ranking functions are used to identify the most relevant documents or passages for a given query~\cite{manning2008ir}.

This retrieval step is a critical component of the pipeline, as it determines the contextual information provided to the generative model. Its effectiveness depends directly on the properties of the similarity measure, which motivates a deeper analysis of the underlying geometric structure.

\subsection{Geometric Tools for Vector Comparison}\label{subsec:geometry}

\subsubsection{Similarity Measures in High-Dimensional Spaces}
A central operation in embedding-based systems is the comparison of vectors in order to assess their semantic proximity. Given two chunk representations $\vec{x}, \vec{y} \in \mathbb{R}^m$, a natural notion of similarity is based on spatial proximity, typically quantified through a distance function.

The most common choice is the Euclidean distance:
\begin{equation}
d(\vec{x},\vec{y}) = \left(\sum_{i=1}^{m}|x_i-y_i|^2\right)^{\frac{1}{2}},
\end{equation}
which measures how close two vectors are in the ambient space, being $x_i$ the components of the numerical vector $\vec{x}$. In the case of $m=2$, the Pythagorean theorem, known from basic geometry is recovered.

In low-dimensional settings, this notion aligns well with intuitive geometric proximity and is therefore a natural candidate for nearest-neighbor retrieval. However, in high-dimensional regimes, the Euclidean distance is sensitive to the scale of the vectors, which may reflect artifacts of the embedding process rather than meaningful semantic differences.

An alternative approach to measuring similarity is based not on proximity, but on the \emph{alignment} between vectors. This perspective relies on the scalar product, defined as:
\begin{equation}
\vec{x}^\top \vec{y} = \sum_{i=1}^m x_i y_i,
\end{equation}
which quantifies the degree to which two vectors point in the same direction. The associated Euclidean norm is given by:
\begin{equation}
\|\vec{x}\| = \sqrt{\vec{x}^\top \vec{x}}.
\end{equation}

The scalar product satisfies the Cauchy-Schwarz inequality:
\begin{equation}
|\vec{x}^\top \vec{y}| \leq \|\vec{x}\| \|\vec{y}\|,
\end{equation}
which naturally leads to a normalized measure of similarity, 
\begin{equation}
-1\leq \dfrac{\vec{x}^\top \vec{y}}{\|\vec{x}\| \|\vec{y}\|} \leq 1,
\end{equation}
namely the cosine of the angle between $\vec{x}$ and $\vec{y}$:
\begin{equation}
\cos(\vec{x},\vec{y}) = \frac{\vec{x}^\top \vec{y}}{\|\vec{x}\| \|\vec{y}\|}.
\end{equation}

Cosine similarity captures the directional agreement between vectors while being invariant to their magnitude/scale. This property makes it particularly well-suited for embedding-based retrieval, where the norm of a vector may encode information unrelated to semantic content (e.g., frequency effects or model-specific scaling).

In practice, modern RAG systems predominantly rely on cosine similarity or closely related metrics for nearest-neighbor search. Nevertheless, as will be shown in the following sections, even angle-based measures are subject to significant degradation in high-dimensional spaces, where both distances and angles may lose discriminative power.

\subsection{Statistical Phenomena in High Dimension}\label{subsec:statistics}
While the above tools behave intuitively in low dimensions, their behavior changes drastically as $m$ increases.

\subsubsection{Distance Concentration Phenomenon}
A fundamental property of high-dimensional spaces is the \emph{concentration of pairwise distances}. This phenomenon is closely related to the broader concentration-of-measure framework developed in high-dimensional probability theory~\cite{ledoux2001concentration}. Let $\vec{x}, \vec{y} \in \mathbb{R}^m$ be two independent random vectors, and consider their Euclidean distance:
\begin{equation}
d(x,y) = \left(\sum_{i=1}^{m}(x_i - y_i)^2\right)^{\frac{1}{2}}.
\end{equation}

Introduce the coordinate-wise deviations $\Delta_i = x_i - y_i$ and define the non-negative contributions: $Z_i = \Delta_i^2.$ The distance can then be rewritten as:
\begin{equation}
d(x,y) = \left(\sum_{i=1}^{m} Z_i \right)^{\frac{1}{2}} = \sqrt{m \, \overline{Z}_m},
\end{equation}
where
\begin{equation}
\overline{Z}_m = \frac{1}{m} \sum_{i=1}^{m} Z_i
\end{equation}
denotes the empirical mean of the coordinate contributions.

Assume that the sequence $(Z_i)_{i=1}^m$ is composed of independent and identically distributed random variables with finite second moment, and define:
\begin{equation}
\mu = \mathbb{E}[Z_i], \qquad \sigma^2 = \mathrm{Var}(Z_i).
\end{equation}

By the Law of Large Numbers,
\begin{equation}
\overline{Z}_m \xrightarrow[m \to \infty]{\mathbb{P}} \mu,
\end{equation}
and by the Central Limit Theorem,
\begin{equation}
\sqrt{m}\,(\overline{Z}_m - \mu) \xrightarrow[m \to \infty]{\mathcal{D}} \mathcal{N}(0, \sigma^2).
\end{equation}

To derive the asymptotic behavior of $d(x,y)$, we apply the delta method to the function $g(s) = \sqrt{m s}$. Since $g'(s) = \frac{\sqrt{m}}{2\sqrt{s}}$, we obtain:
\begin{equation}
\mathrm{Var}(d(x,y)) \approx \frac{\sigma^2}{4\mu}.
\end{equation}

In particular, the variance remains bounded as $m$ increases, while the mean behaves as:
\begin{equation}
\mathbb{E}[d(x,y)] \approx \sqrt{m \mu}.
\end{equation}

This yields the following approximation for the coefficient of variation:
\begin{equation}
\mathrm{CV}(d(x,y)) = \frac{\sqrt{\mathrm{Var}(d(x,y))}}{\mathbb{E}[d(x,y)]}
\approx \frac{\sigma}{2\mu} \cdot \frac{1}{\sqrt{m}}.
\end{equation}

Therefore,
\begin{equation}
\mathrm{CV}(d(x,y)) \xrightarrow[m \to \infty]{} 0,
\end{equation}
which implies that pairwise distances become increasingly concentrated around their mean. The concentration of pairwise distances observed here is a standard manifestation of high-dimensional probability phenomena~\cite{vershynin2018high}.

A complementary measure of concentration is the \emph{relative contrast}, which quantifies the spread of distances from a given query point to a finite dataset. Let $\vec{x} \in \mathbb{R}^m$ be a query vector and let $\mathcal{X} = \{\vec{x}_1, \vec{x}_2, \ldots, \vec{x}_N\} \subset \mathbb{R}^m$ denote a collection of points. Define:
\begin{align}
d_{\min}(\vec{x}) &= \min_{1 \leq i \leq N} \|\vec{x} - \vec{x}_i\|, \\
d_{\max}(\vec{x}) &= \max_{1 \leq i \leq N} \|\vec{x} - \vec{x}_i\|.
\end{align}

The relative contrast is then given by:
\begin{equation}
\mathrm{RC}(\vec{x}) = \frac{d_{\max}(\vec{x}) - d_{\min}(\vec{x})}{d_{\min}(\vec{x})}.
\end{equation}

This quantity measures how distinguishable the nearest neighbor is from the farthest point in the dataset. As the dimension increases, $\mathrm{RC}(\vec{x})$ typically decreases, indicating that all distances become comparable and that nearest-neighbor distinctions lose significance. This loss of contrast is a key mechanism underlying the degradation of similarity-based retrieval in high-dimensional spaces.

\subsubsection{Asymptotic Orthogonality}
A second manifestation of high-dimensional concentration appears at the angular level. Let $\vec{x},\vec{y} \in \mathbb{R}^m$ be two independent random vectors with independent coordinates, zero mean, finite variance, and finite fourth moments. For simplicity, assume
\begin{equation}
\mathbb{E}[x_i] = \mathbb{E}[y_i] = 0,
\qquad
\mathbb{E}[x_i^2] = \mathbb{E}[y_i^2] = \sigma_x^2,
\end{equation}
with analogous assumptions for $\vec{y}$.

The scalar product can be written as
\begin{equation}
\vec{x}^\top \vec{y} = \sum_{i=1}^m x_i y_i.
\end{equation}

Since $\vec{x}$ and $\vec{y}$ are independent and centered,
\begin{equation}
\mathbb{E}[x_i y_i] = 0.
\end{equation}

Moreover,
\begin{equation}
\mathrm{Var}(\vec{x}^\top \vec{y}) = \sum_{i=1}^m \mathrm{Var}(x_i y_i) = m \, \sigma_x^2 \sigma_y^2.
\end{equation}

Hence the numerator of the cosine similarity fluctuates at scale $\sqrt{m}$. On the other hand, by the Law of Large Numbers,
\begin{equation}
\frac{1}{m}\|x\|^2 = \frac{1}{m}\sum_{i=1}^m x_i^2 \xrightarrow[m\to\infty]{\mathbb{P}}
\sigma_x^2, 
\end{equation}
and similarly
\begin{equation}
\frac{1}{m}\|y\|^2 \xrightarrow[m\to\infty]{\mathbb{P}} \sigma_y^2.
\end{equation}

Therefore,
\begin{equation}
\|x\|\|y\| \approx m \sigma_x \sigma_y.
\end{equation}

Combining the scaling of the numerator and denominator gives
\begin{equation}
\cos(x,y) = \frac{\vec{x}^\top \vec{y}}{\|\vec{x}\|\|\vec{y}\|} = O_{\mathbb{P}}\left(\frac{1}{\sqrt{m}}\right).
\end{equation}

Consequently,
\begin{equation}
\cos(x,y) \xrightarrow[m\to\infty]{\mathbb{P}} 0.
\end{equation}
Such asymptotic orthogonality properties are characteristic of random vectors in high-dimensional spaces~\cite{vershynin2018high}.

This result expresses the phenomenon of \emph{asymptotic orthogonality}: in high dimension, two independent centered random vectors become nearly orthogonal with high probability. More precisely, their angular similarity concentrates around zero, with fluctuations of order $m^{-1/2}$. 

This observation is important for embedding-based systems because it shows that, once dominant mean effects are removed, random high-dimensional vectors tend to exhibit very weak angular separation. As a result, cosine similarity values may become increasingly difficult to distinguish from one another.

\subsubsection{Retrieval Instability}
The previous results have direct consequences for nearest-neighbor retrieval. Consider a query vector $\vec{x} \in \mathbb{R}^m$ and a finite collection of candidate embeddings
\begin{equation}
\mathcal{X} = \{\vec{x}_1,\vec{x}_2,\ldots,\vec{x}_N\} \subset \mathbb{R}^m.
\end{equation}

For cosine-based retrieval, each candidate is assigned the score
\begin{equation}
s_i = \cos(\vec{x},\vec{x}_i).
\end{equation}

The retrieved elements are obtained by sorting these scores. Let $\vec{x}_{(1)}$ and $\vec{x}_{(2)}$ denote the first and second retrieved neighbors, so that
\begin{equation}
s_{(1)} = \cos(\vec{x},\vec{x}_{(1)}), \; s_{(2)} = \cos(\vec{x},\vec{x}_{(2)}), \; s_{(1)} \geq s_{(2)}.
\end{equation}

The \emph{cosine gap} is defined as
\begin{equation}
\Delta(\vec{x}) = s_{(1)} - s_{(2)} = \cos(\vec{x},\vec{x}_{(1)}) - \cos(\vec{x},\vec{x}_{(2)}).
\end{equation}

This quantity measures the separation between the best and second-best retrieval candidates. A large gap indicates a stable retrieval decision, whereas a small gap indicates ambiguity.

Under the asymptotic orthogonality regime, the scores $s_i$ concentrate around zero. In simplified isotropic settings, one may interpret
\begin{equation}
s_i = O_{\mathbb{P}}\left(\frac{1}{\sqrt{m}}\right).
\end{equation}

Therefore, the separation between competing cosine scores also shrinks with dimension:
\begin{equation}
\Delta(\vec{x}) \to 0 \qquad \text{as} \qquad m \to \infty,
\end{equation}
up to factors depending on the number of candidates $N$ and their dependence structure.

This shrinking gap produces retrieval instability. Indeed, suppose the query or candidate embeddings are perturbed by a small error term, for example due to numerical noise, embedding model variability, chunking choices, or semantic ambiguity:
\begin{equation}
\vec{x} \longrightarrow \vec{x} + \vec{\varepsilon}.
\end{equation}

If the perturbation changes the retrieval scores by an amount comparable to or larger than $\Delta(\vec{x})$, then the ordering of the nearest neighbors may change:
\begin{equation}
|\delta s_1 - \delta s_2| > \Delta(\vec{x}) \quad \Longrightarrow \quad \vec{x}_{(2)} \text{ may replace } \vec{x}_{(1)}.
\end{equation}

Thus, as the cosine gap decreases, increasingly small perturbations become sufficient to alter the retrieved context.

In RAG systems, this mechanism is operationally significant. The generative model does not receive the entire knowledge base, but only the retrieved subset of chunks. If high-dimensional concentration makes retrieval ambiguous, then the supplied context may become unstable, incomplete, or weakly relevant. This provides a geometric pathway through which embedding-space concentration can contribute to weak grounding and, consequently, increase hallucination risk.

\section{Numerical Experiments}\label{sec:experiments}
The objective of this section is to empirically characterize the geometric and statistical phenomena described in the previous sections, and to quantify their practical impact on similarity-based retrieval. Rather than relying on a specific embedding model, we adopt a controlled experimental framework based on synthetic data. This allows us to isolate the intrinsic effects of high-dimensional geometry from model-dependent artifacts such as training bias, anisotropy, or dataset-specific structure.

\subsection{Experimental Protocol}\label{subsec:protocol}
We consider synthetic datasets $\mathcal{X} = \{\vec{x}_1,\vec{x}_2, \ldots, \vec{x}_N\} \subset \mathbb{R}^m$ with a fixed sample size $N = 500$, while the ambient dimension $m$ varies over a wide range, from low-dimensional regimes ($m=2$) up to high-dimensional settings ($m=10^4$), with logarithmic spacing to capture both transitional and asymptotic behaviors. The samples are drawn independently from a collection of probability distributions chosen to span a broad spectrum of statistical structures, including light-tailed, heavy-tailed, skewed, and sparse regimes. Specifically, we consider Gaussian distributions $\mathcal{N}(0,1)$, uniform distributions $\mathcal{U}(-1,1)$, Poisson distributions with parameter $\lambda=3$, Beta distributions $\mathrm{Beta}(2,5)$, lognormal distributions, Student-$t$ distributions to capture heavy-tailed behavior, and sparse Bernoulli distributions with success probability $p=0.05$.

In order to disentangle distributional effects from intrinsic geometric phenomena, we analyze two complementary preprocessing regimes. In the raw setting, the samples are used as generated, preserving their original statistical structure, including potential mean shifts and anisotropy. In the standardized setting, each coordinate is centered and rescaled to unit variance, thereby removing first- and second-order effects and isolating the geometry induced by high dimensionality itself. This distinction is essential, as real-world embedding spaces often exhibit non-centered and anisotropic behavior, which may artificially inflate similarity measures.

For each configuration, we compute a set of geometric and statistical observables designed to capture different aspects of high-dimensional behavior. These include the mean and standard deviation of pairwise cosine similarity, which characterize angular concentration; the coefficient of variation of pairwise distances, which quantifies distance concentration; and the relative contrast, which measures the separation between nearest and farthest neighbors. To assess retrieval-specific effects, we also evaluate cosine score gaps between top-ranked neighbors, stability metrics based on Jaccard similarity under perturbations, and top-1 change rates, which directly quantify ranking instability. In addition, we analyze hubness through inequality-based measures such as the Gini coefficient and distributional skewness, capturing the emergence of points that appear disproportionately often in nearest-neighbor lists.

Finally, we complement these geometric diagnostics with a simplified simulation of a RAG setting. In this framework, we evaluate retrieval success rates and derive a proxy for hallucination risk based on the failure to retrieve relevant neighbors. Although the datasets are synthetic, each metric has a direct interpretation in production systems: cosine similarity underlies embedding search engines, distance concentration impacts vector database discriminability, score gaps reflect ranking confidence, stability metrics capture retrieval robustness, and hubness indicates structural bias in the representation space. As such, the experimental setup can be interpreted as a controlled stress test of embedding-based retrieval pipelines, providing insights that extend beyond the specific models considered here.

\subsection{Cosine Similarity Concentration}
\begin{figure}[htbp]
\centering
\includegraphics[width=0.45\textwidth]{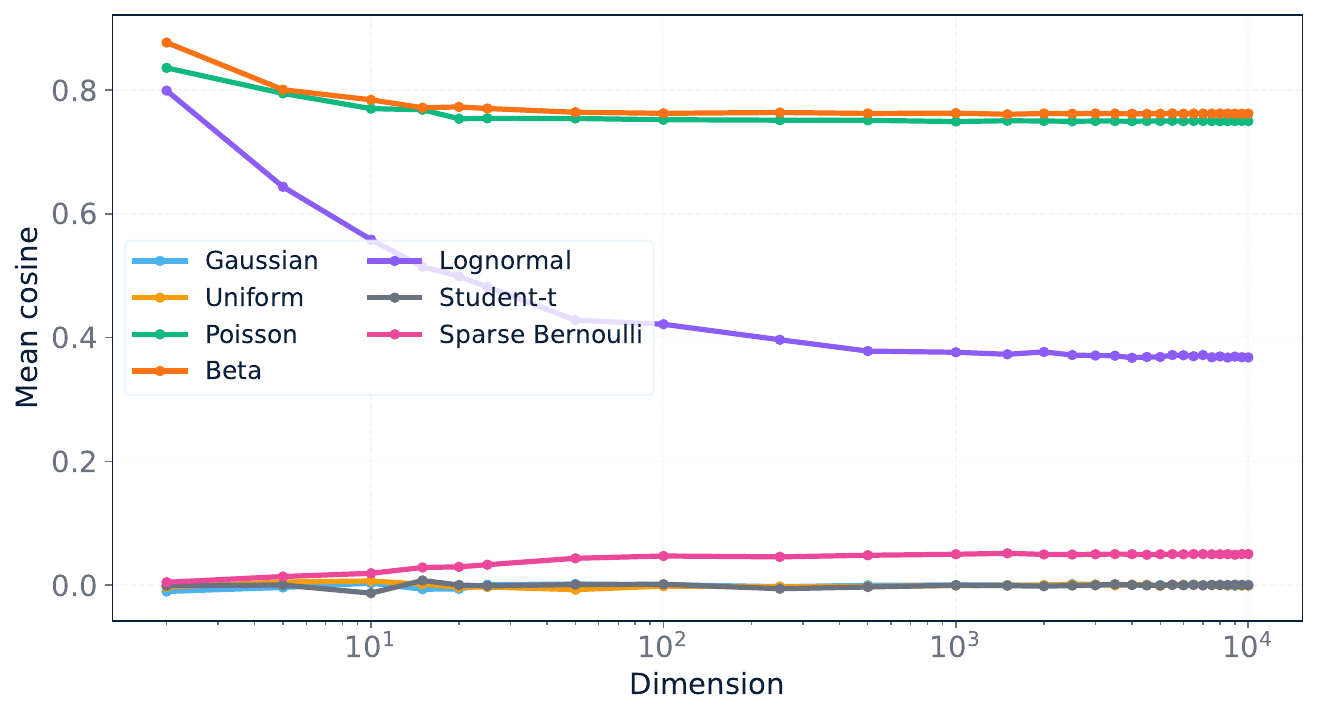}\\
\includegraphics[width=0.45\textwidth]{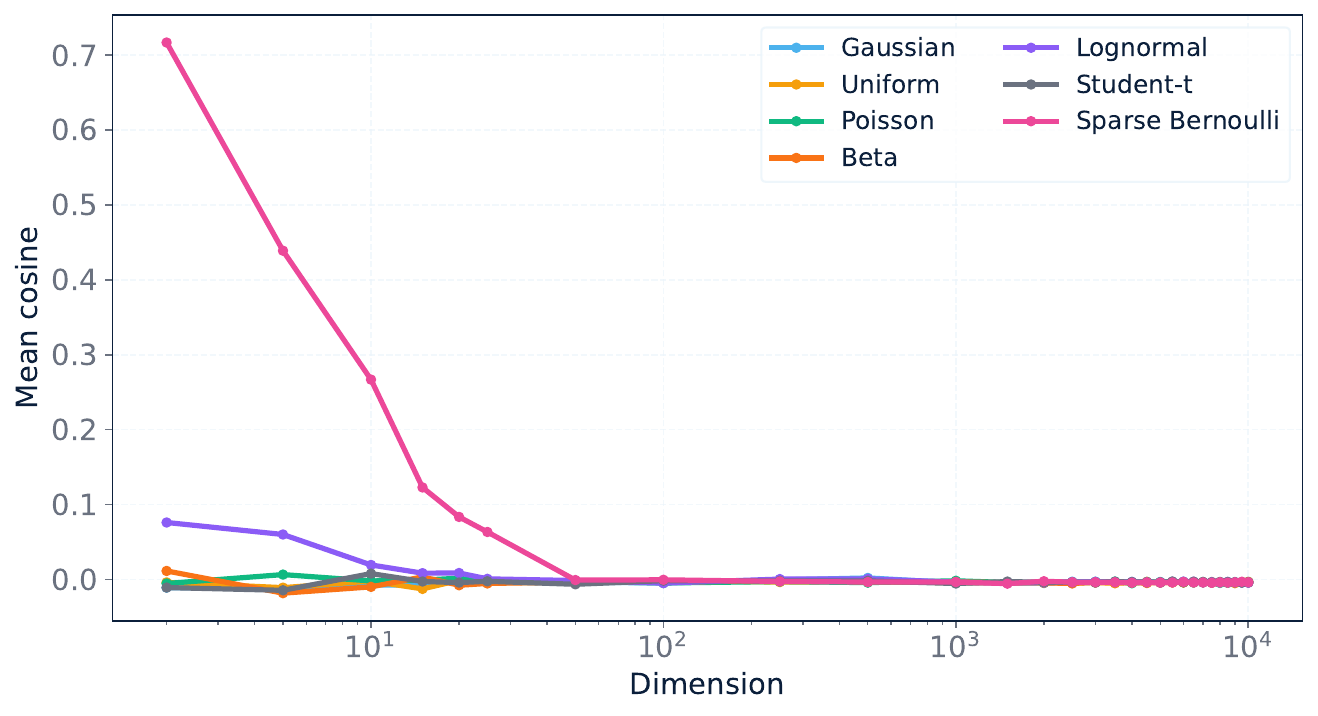}
\caption{(color online) Mean pairwise cosine similarity as a function of dimension. Top panel: raw space. Bottom panel: standardized space.}
\label{fig:cosine_combined}
\end{figure}

The behavior of cosine similarity in high-dimensional spaces exhibits two distinct regimes, depending on whether distributional biases are present. These regimes are illustrated in Fig.~\ref{fig:cosine_combined}, where the left panel corresponds to raw data and the right panel to standardized data.

In the raw space (top panel), cosine similarity strongly depends on the underlying data distribution. In particular, distributions supported on the positive orthant, such as Poisson, Beta, and Lognormal, exhibit a high level of alignment, with mean cosine values stabilizing at significantly positive levels even as the dimension increases. In contrast, centered distributions such as Gaussian, Uniform, and Student-$t$ rapidly converge toward zero.

This behavior can be explained through the decomposition of the scalar product. For non-centered distributions, one has
\begin{equation}
\mathbb{E}[\vec{x}^\top \vec{y}] = m \, \mathbb{E}[\vec{x}_i] \, \mathbb{E}[\vec{y}_i] \neq 0,
\end{equation}
which induces a systematic positive bias in cosine similarity. As a result, independently sampled vectors are not isotropically distributed around the origin but instead exhibit a preferred direction, leading to artificially high similarity scores. In this regime, cosine similarity is therefore dominated by first-order statistical moments rather than intrinsic geometric relationships.

After standardization (bottom panel), this bias is removed and all distributions collapse toward a common behavior. In this setting, the data is centered so that $\mathbb{E}[\vec{x}_i] = 0$, implying
\begin{equation}
\mathbb{E}[\vec{x}^\top \vec{y}] = 0,
\end{equation}
and leaving only fluctuations around zero. As shown in Section~\ref{subsec:statistics}, the scalar product scales as $O(\sqrt{m})$ while the norms scale as $O(m)$, yielding
\begin{equation}
\cos(x,y) = O_{\mathbb{P}}\left(\frac{1}{\sqrt{m}}\right),
\end{equation}
and therefore
\begin{equation}
\cos(x,y) \xrightarrow[m\to\infty]{} 0,
\end{equation}
with vanishing variance. This confirms the phenomenon of asymptotic orthogonality.

Empirically, the transition toward this concentration regime occurs beyond a critical dimension $m^*$, which can be interpreted as the scale at which high-dimensional geometry dominates over distribution-specific structure. In the present experiments, this transition is observed around $m^* \simeq 10^2$, beyond which all distributions exhibit nearly identical cosine behavior.

Taken together, these results reveal a two-stage mechanism governing cosine similarity in high-dimensional spaces. In raw representations, similarity is dominated by distributional biases such as mean shifts, sparsity, or skewness, leading to artificially inflated alignment. After removing these biases through standardization, the intrinsic geometric structure emerges, but this structure is itself degenerate: cosine values concentrate near zero and become increasingly indistinguishable.

From an application perspective, this dual phenomenon has important implications for embedding-based retrieval systems. In raw embedding spaces, spurious similarity may arise due to anisotropy or dominant directions, causing unrelated items to appear artificially close. In contrast, in normalized or centered spaces, the concentration of cosine values reduces the contrast between candidates, making ranking decisions more sensitive to perturbations and noise. Related anisotropy effects have also been observed empirically in contextualized word representations produced by transformer-based language models~\cite{ethayarajh2019anisotropy}.

Therefore, cosine similarity suffers from a fundamental trade-off in high-dimensional settings: removing bias exposes concentration, while preserving structure risks introducing spurious alignment. This observation motivates the need for additional diagnostics, such as variance analysis, score gaps, and stability metrics, which are investigated in the following subsections.

\subsection{Variance of Cosine Similarity}
\begin{figure}[htbp]
\centering
\includegraphics[width=0.45\textwidth]{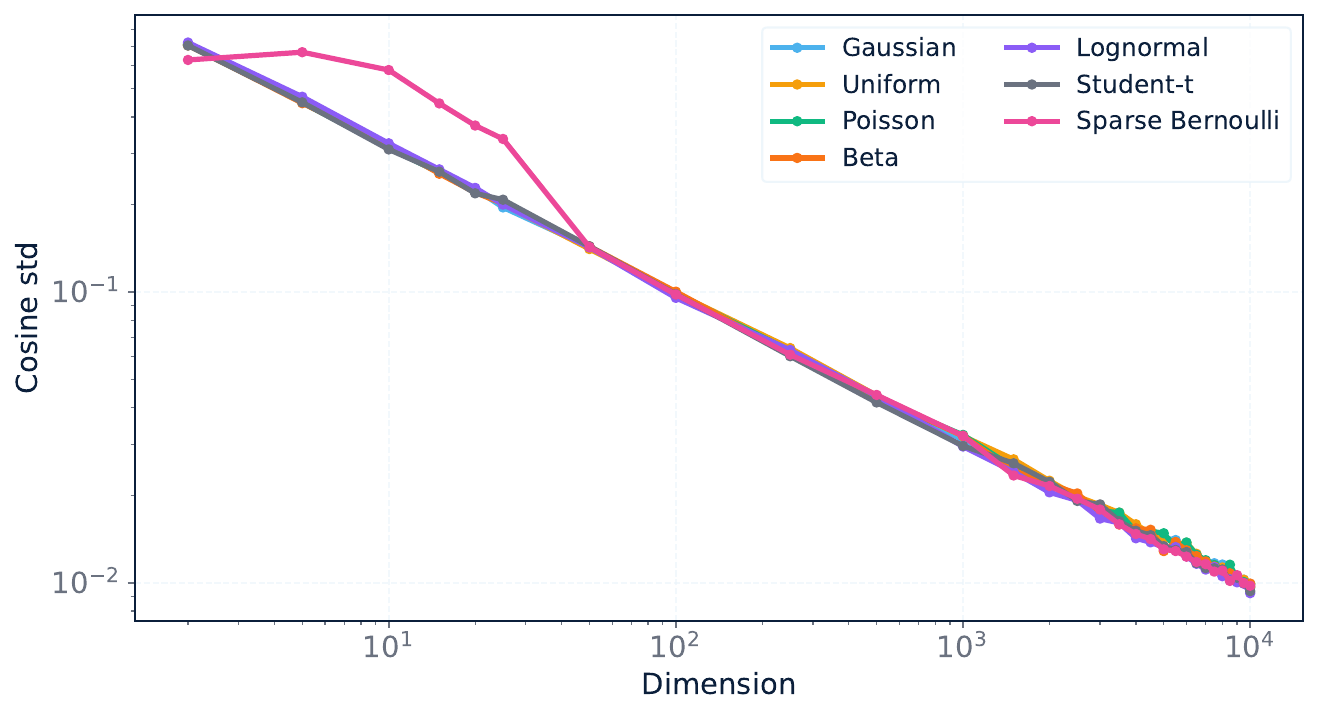}
\caption{(color online) Standard deviation of cosine similarity as a function of dimension.}
\label{fig:cosine_var}
\end{figure}

Beyond the convergence of the mean cosine similarity to zero, an equally important effect is the collapse of its variability. Fig.~\ref{fig:cosine_var} shows that the standard deviation of pairwise cosine similarity decreases approximately as $m^{-1/2}$, in agreement with the theoretical scaling derived in Section~\ref{subsec:statistics}.

This behavior reflects a stronger form of concentration than that observed at the level of the mean. Indeed, while Fig.~\ref{fig:cosine_combined} showed that cosine values converge toward zero after standardization, Fig.~\ref{fig:cosine_var} demonstrates that their dispersion simultaneously vanishes. In other words, not only do most vectors become nearly orthogonal, but they also become \emph{uniformly} orthogonal, with increasingly negligible variability across pairs.

From a probabilistic perspective, this implies that the distribution of cosine similarity collapses toward a Dirac mass at zero. More precisely, for independent centered vectors, one has
\begin{equation}
\cos(x,y) = O_{\mathbb{P}}\left(\frac{1}{\sqrt{m}}\right),
\end{equation}
so that both the expectation and the standard deviation decay at the same rate. As a consequence, typical fluctuations of cosine similarity become indistinguishable from zero beyond moderate dimensions.

Empirically, this effect manifests at dimensions comparable to those encountered in modern embedding systems. As observed in the previous subsection, the transition toward this regime occurs around a critical scale $m^* \simeq 10^2$, beyond which the spread of cosine values becomes extremely narrow and nearly identical across all distributions.

This collapse of variance has critical implications for similarity-based retrieval. In practical systems, cosine similarity is used as a ranking signal under the assumption that higher values indicate stronger semantic relevance. However, when the variance shrinks, the distribution of scores becomes highly concentrated, and the differences between candidates fall within a narrow band. As a result, even small perturbations (arising from numerical noise, embedding variability, or minor semantic differences) may produce changes of the same order as the intrinsic variability of the scores.

Consequently, the ranking induced by cosine similarity becomes increasingly unstable, as it relies on comparing quantities that are nearly indistinguishable. This effect provides a quantitative explanation for the degradation of retrieval reliability in high-dimensional embedding spaces, and directly motivates the analysis of score gaps and stability metrics in the following subsections.

\subsection{Distance Concentration}
\begin{figure}[htbp]
\centering
\includegraphics[width=0.45\textwidth]{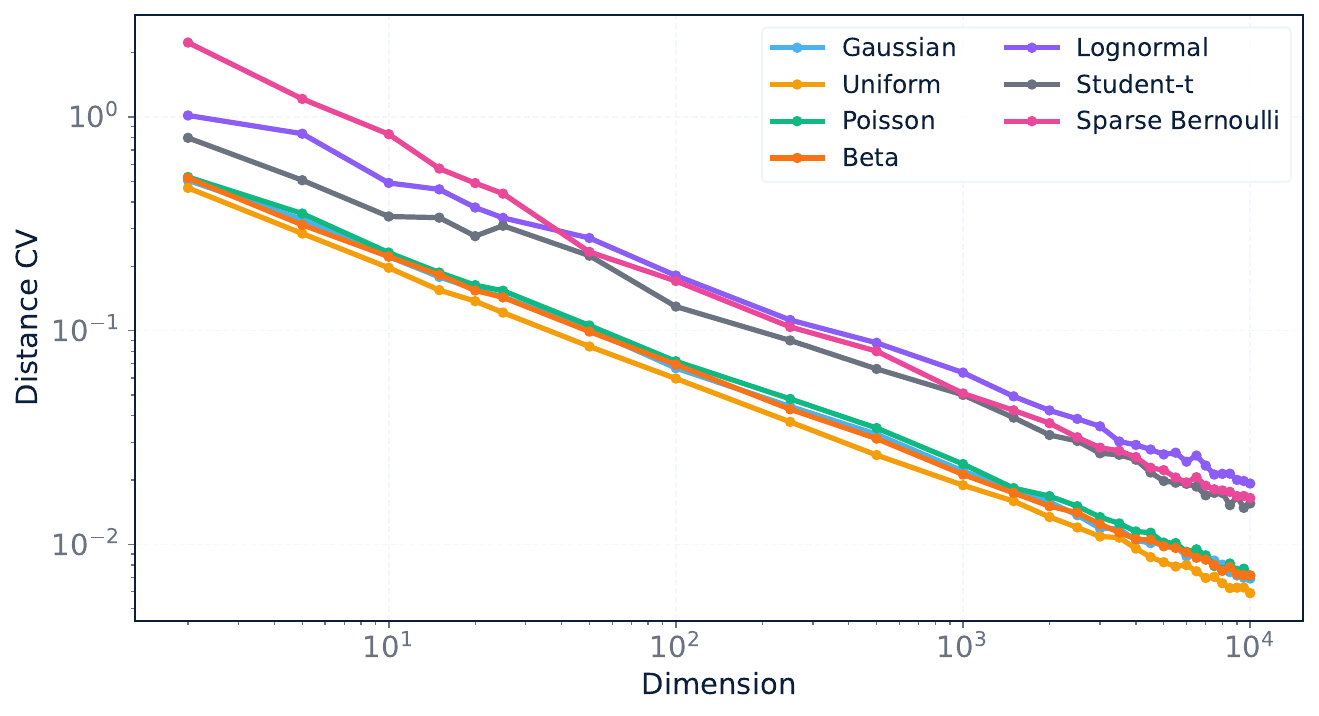}
\caption{(color online) Coefficient of variation of pairwise distances as a function of dimension.}
\label{fig:distance_cv}
\end{figure}

A complementary manifestation of high-dimensional effects appears at the level of Euclidean distances. Fig.~\ref{fig:distance_cv} shows that the coefficient of variation of pairwise distances decreases sharply with the dimension, in agreement with the theoretical scaling derived in Section~\ref{subsec:statistics}.

More precisely, the coefficient of variation behaves as
\begin{equation}
\mathrm{CV}(d(x,y)) \approx \frac{\sigma}{2\mu}\,\frac{1}{\sqrt{m}},
\end{equation}
so that
\begin{equation}
\mathrm{CV}(d(x,y)) \xrightarrow[m\to\infty]{} 0.
\end{equation}

This implies that pairwise distances concentrate around their mean value, with fluctuations that become negligible relative to the scale of the distances themselves. In probabilistic terms, the distribution of distances collapses toward a narrow band centered around $\mathbb{E}[d(x,y)]$, reflecting a concentration-of-measure phenomenon.

Importantly, this effect is remarkably robust across all considered distributions. As shown in Fig.~\ref{fig:distance_cv}, once the data is standardized, the curves corresponding to Gaussian, Uniform, heavy-tailed, and sparse distributions all exhibit the same decay pattern, indicating that distance concentration is an intrinsic property of high-dimensional geometry rather than a consequence of specific statistical assumptions.

From a geometric perspective, this result implies that the notion of proximity becomes increasingly degenerate: the ratio between the distance to the nearest neighbor and the distance to a typical point approaches unity. In other words, the distinction between ``close'' and ``far'' points progressively vanishes as the dimension increases.

This has direct implications for nearest-neighbor search and vector-based retrieval systems. In low-dimensional settings, Euclidean distance provides a meaningful ordering of points based on proximity. However, in high dimensions, the collapse of variability implies that many points lie at comparable distances from a given query, reducing the discriminative power of distance-based ranking.

From an operational standpoint, this phenomenon translates into a degradation of retrieval quality. Since candidate points are no longer well separated in terms of distance, small perturbations (arising from numerical precision, embedding noise, or data variability) can significantly alter the ordering of neighbors. As a consequence, distance-based retrieval becomes less stable and less reliable.

Taken together with the cosine concentration results, this analysis reveals a fundamental limitation of similarity measures in high-dimensional spaces: both distance and angular metrics suffer from concentration effects that reduce contrast and impair discrimination. This observation motivates the need for alternative diagnostics and robustness measures, which are explored in the subsequent subsections.

\subsection{Relative Contrast}
\begin{figure}[htbp]
\centering
\includegraphics[width=0.45\textwidth]{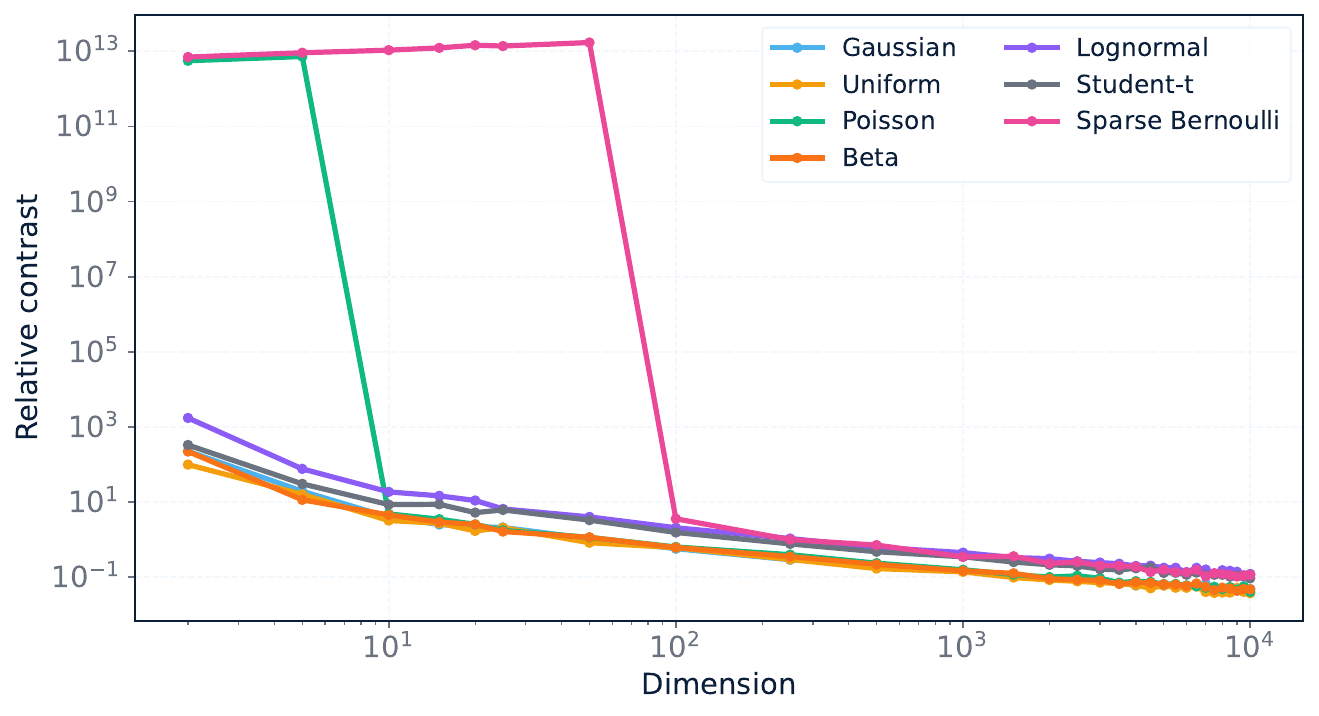}
\caption{(color online) Relative contrast as a function of dimension.}
\label{fig:rc}
\end{figure}

A complementary and more operational measure of distance concentration is provided by the relative contrast. As shown in Fig.~\ref{fig:rc}, the relative contrast decreases rapidly with dimension, indicating that the gap between the nearest and farthest neighbors collapses.

Recall that, for a query $\vec{x} \in \mathbb{R}^m$ and a dataset $\mathcal{X}$, the relative contrast is defined as
\begin{equation}
\mathrm{RC}(\vec{x}) = \frac{d_{\max}(\vec{x}) - d_{\min}(\vec{x})}{d_{\min}(\vec{x})},
\end{equation}
where $d_{\min}(\vec{x})$ and $d_{\max}(\vec{x})$ denote the distances from $\vec{x}$ to its nearest and farthest neighbors, respectively. This quantity directly measures the separation between the most relevant candidate and the rest of the dataset.

The rapid decay observed in Fig.~\ref{fig:rc} shows that, as the dimension increases, $d_{\max}(\vec{x})$ and $d_{\min}(\vec{x})$ become increasingly similar, so that
\begin{equation}
\mathrm{RC}(\vec{x}) \xrightarrow[m\to\infty]{} 0.
\end{equation}

This result refines the conclusions drawn from the coefficient of variation. While the latter captures the global concentration of pairwise distances, relative contrast focuses on the \emph{extreme values} that govern nearest-neighbor retrieval. The collapse of $\mathrm{RC}(\vec{x})$ therefore implies that even the most favorable comparison (between the closest and farthest points) loses discriminative power.

From a geometric standpoint, this means that the metric structure of the space becomes increasingly flat: all points lie at comparable distances from the query, and the notion of a well-defined neighborhood progressively disappears. In such a regime, nearest-neighbor search becomes intrinsically ill-posed, as the ordering of candidates is no longer supported by meaningful geometric separation.

This effect is particularly critical in embedding-based retrieval systems. In practical RAG pipelines, the relevance of retrieved chunks is determined by selecting the nearest neighbors of a query. However, when the relative contrast collapses, the difference between relevant and irrelevant candidates becomes negligible, and retrieval decisions become highly sensitive to perturbations, noise, or modeling artifacts.

Therefore, relative contrast provides a direct and interpretable measure of the loss of retrieval discriminability in high-dimensional spaces. Combined with the previous results on cosine and distance concentration, it highlights a fundamental limitation: even under idealized conditions, the geometric signal used to drive retrieval may become insufficient to support robust decision-making. This observation motivates the analysis of score gaps and stability metrics, which capture how these geometric effects translate into practical retrieval failures.

\subsection{Cosine Gap and Ranking Confidence}
\begin{figure}[htbp]
\centering
\includegraphics[width=0.45\textwidth]{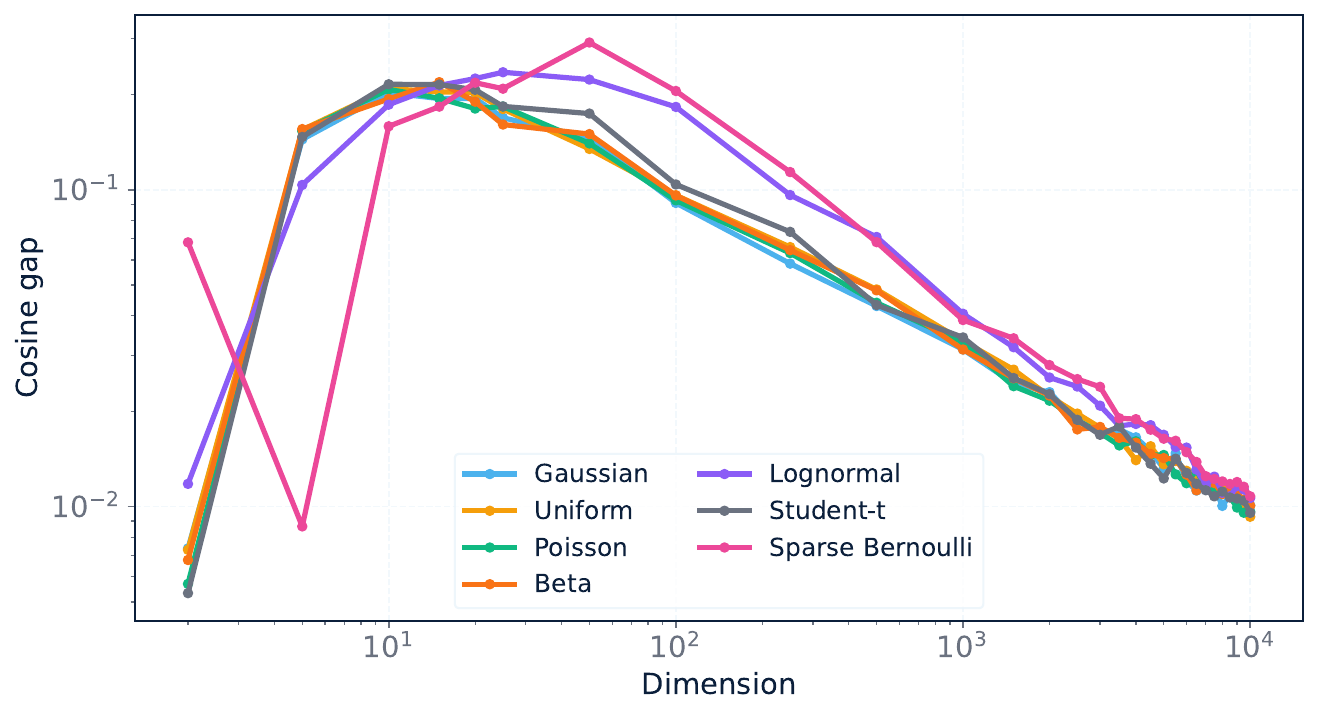}
\caption{(color online) Cosine gap $\Delta(\vec{x})$ between top-ranked neighbors as a function of dimension.}
\label{fig:gap}
\end{figure}

A direct operational consequence of cosine concentration is the collapse of score separation between competing candidates. As shown in Fig.~\ref{fig:gap}, the cosine gap $\Delta(\vec{x})$ decreases steadily with the dimension, indicating that the difference between the best and subsequent retrieval candidates becomes progressively smaller.

The gap $\Delta(\vec{x})$ measures the \emph{confidence margin} of the retrieval decision: a large gap indicates a clear distinction between the top candidate and its competitors, whereas a small gap signals ambiguity. In the experiments, we extend this notion to broader comparisons (e.g., top-1 versus top-$k$), capturing the effective separation between the leading candidate and the bulk of the distribution.

The observed decay of $\Delta(\vec{x})$ with the embedding dimension is consistent with the scaling of cosine similarity. As shown in the previous subsections, cosine values concentrate around zero with fluctuations of order $O(m^{-1/2})$. Consequently, the differences between top-ranked scores also shrink at a comparable rate:
\begin{equation}
\Delta(\vec{x}) = O_{\mathbb{P}}\left(\frac{1}{\sqrt{m}}\right),
\end{equation}
which implies
\begin{equation}
\Delta(\vec{x}) \xrightarrow[m\to\infty]{} 0.
\end{equation}

From a statistical perspective, this collapse of the gap indicates that the signal used to discriminate between candidates vanishes relative to the scale of fluctuations. In other words, the ranking is no longer supported by a robust separation in similarity scores, but instead relies on increasingly small differences that are sensitive to sampling variability.

This phenomenon has direct implications for retrieval systems. In practice, cosine similarity is used to order candidates under the assumption that higher scores correspond to more relevant items. However, when the cosine gap becomes small, even minor perturbations (arising from numerical precision, stochastic embedding generation, or slight variations in input text) may exceed the intrinsic separation between candidates and lead to changes in the ranking.

Importantly, this ambiguity arises even in the absence of noise: it is a structural consequence of high-dimensional concentration. Thus, the collapse of the cosine gap provides a quantitative bridge between geometric effects and retrieval uncertainty. It explains why nearest-neighbor selection becomes unreliable and motivates the need to analyze stability metrics, which explicitly measure the sensitivity of retrieval outcomes to perturbations in the embedding space.

\subsection{Hubness Phenomenon}
\begin{figure}[htbp]
\centering
\includegraphics[width=0.45\textwidth]{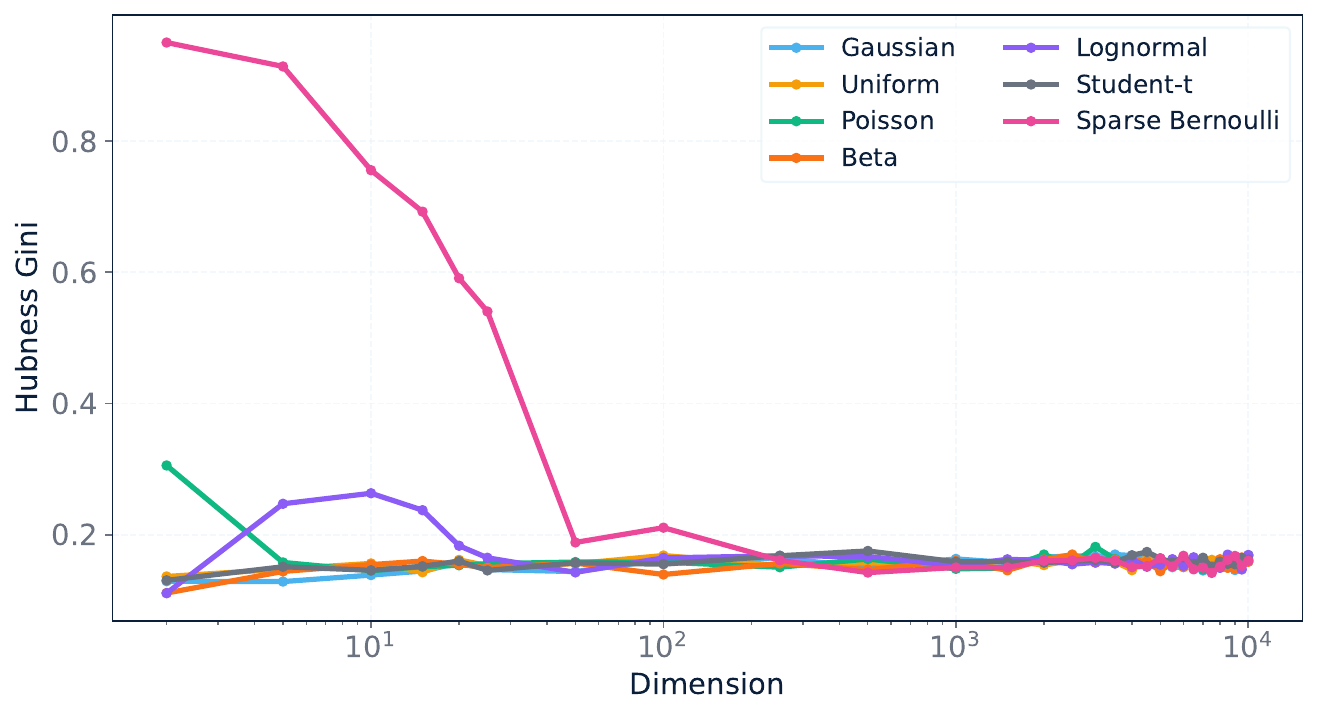}\\
\includegraphics[width=0.45\textwidth]{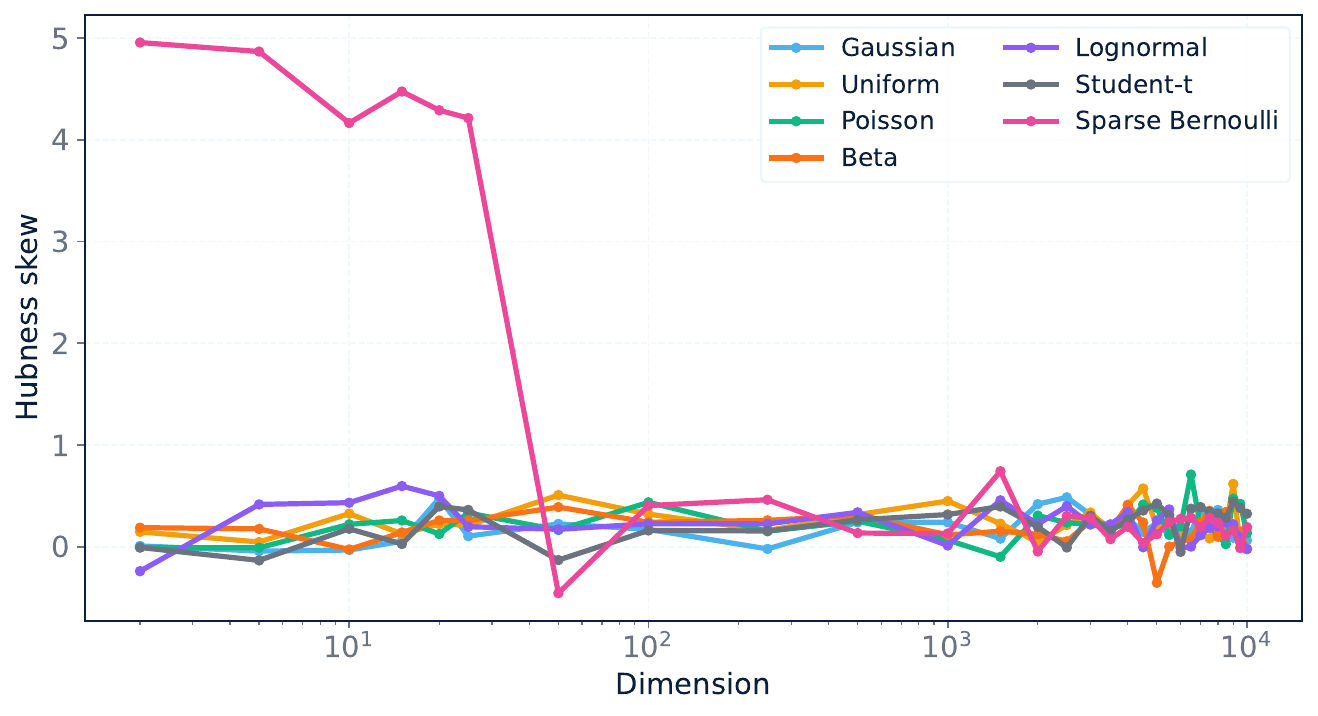}
\caption{(color online) Hubness metrics as a function of dimension. Top panel: Gini coefficient. Bottom panel: skewness.}
\label{fig:hubness}
\end{figure}

A further consequence of high-dimensional concentration is the emergence of \emph{hubness}, a structural bias in which certain points appear disproportionately often in nearest-neighbor lists. This phenomenon is illustrated in Fig.~\ref{fig:hubness}, where both the Gini coefficient (top panel) and the skewness of neighbor occurrence distributions (bottom panel) increase with the dimension.

These results indicate that nearest-neighbor relationships become increasingly asymmetric: while most points are rarely selected, a small subset of points (the so-called hubs) are retrieved with high frequency across many queries. Importantly, this effect is not tied to any specific distribution, but arises generically as a consequence of high-dimensional geometry.

From a geometric standpoint, hubness reflects the uneven distribution of points with respect to the global structure of the space. Certain vectors, by virtue of their position relative to the bulk of the data, become centrally located and thus tend to minimize distance or maximize similarity to many other points simultaneously. As dimensionality increases, this imbalance is amplified, leading to a highly skewed retrieval landscape.

This phenomenon has important implications for embedding-based systems. In retrieval pipelines, hub vectors may act as attractors, appearing in the results of many unrelated queries. As a consequence, they can introduce systematic bias, reduce diversity in retrieved content, and increase the likelihood of irrelevant or generic information being selected.

In the context of RAG systems, hubness can contribute to the propagation of spurious context. Since certain chunks are retrieved disproportionately often, they may dominate the input to the generative model, even when they are only weakly related to the query. This can degrade the quality of generated outputs and amplify misleading or irrelevant information.

Therefore, hubness provides a complementary perspective on the limitations of similarity-based retrieval in high-dimensional spaces. While previous subsections highlighted the loss of contrast and the collapse of score differences, the present analysis reveals an additional structural effect: the emergence of biased retrieval patterns driven by the geometry of the embedding space itself.

\subsection{Retrieval Instability and Impact on RAG Performance}
\begin{figure}[htbp]
\centering
\includegraphics[width=0.45\textwidth]{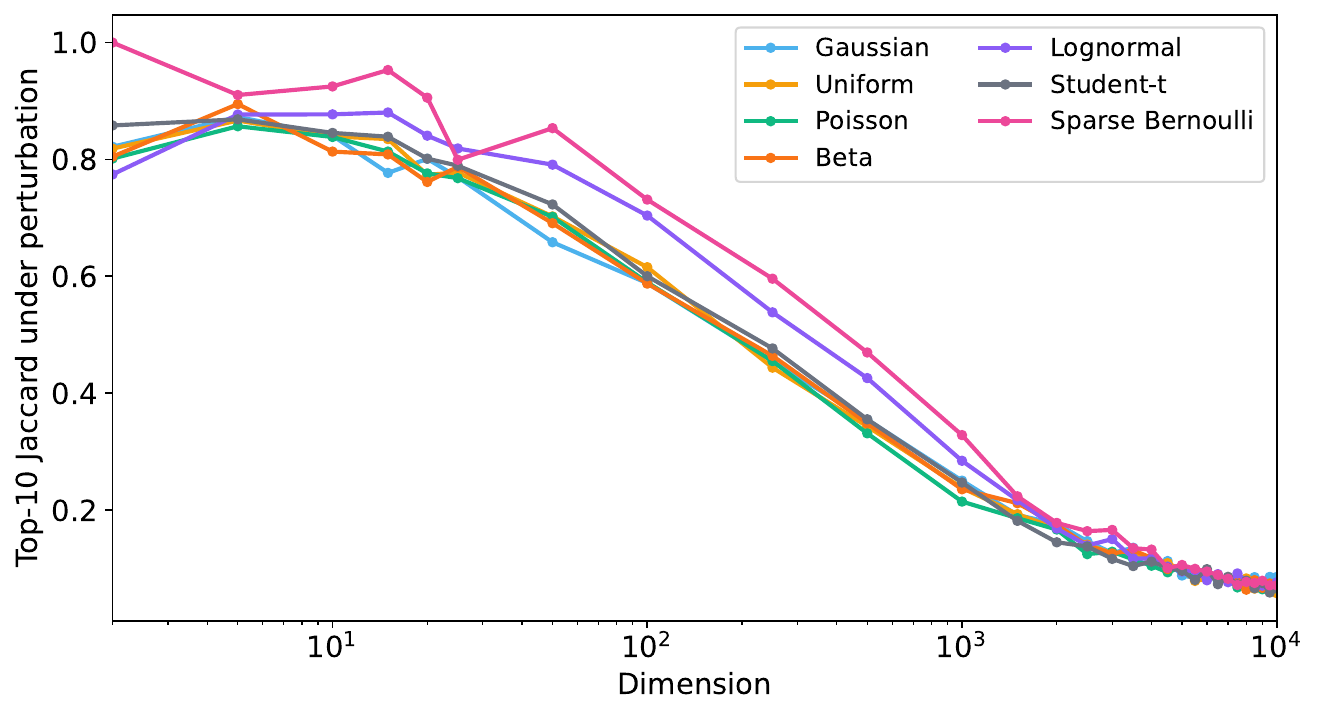}\\
\includegraphics[width=0.45\textwidth]{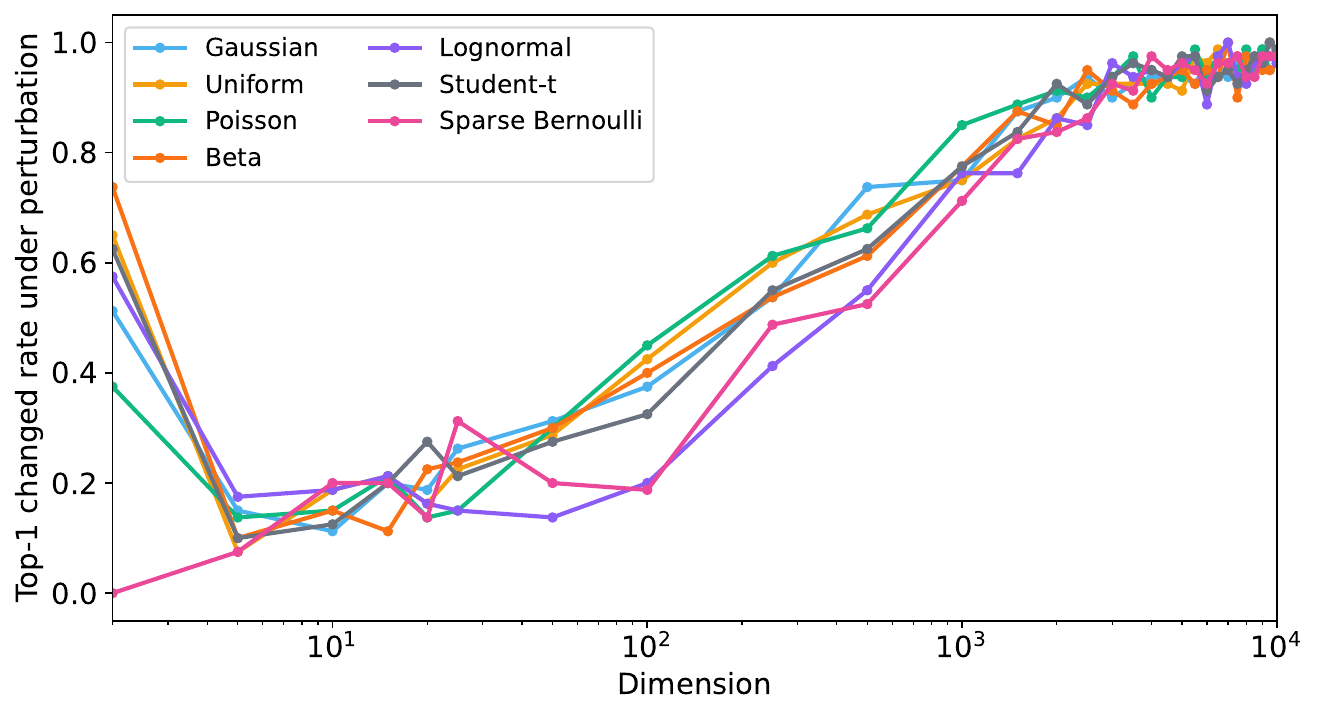}
\caption{(color online) Retrieval stability under perturbations. Top panel: Jaccard similarity of top-$k$ sets. Bottom panel: top-1 change rate.}
\label{fig:retstability}
\end{figure}

The geometric effects described in the previous subsections translate directly into observable instability in retrieval systems. Fig.~\ref{fig:retstability} illustrates this phenomenon through two complementary metrics: the Jaccard similarity between retrieved sets under small perturbations (top panel) and the rate at which the top-1 retrieved element changes (bottom panel).

As the dimension increases, the Jaccard similarity decreases sharply, indicating that the overlap between retrieved sets becomes increasingly unstable. At the same time, the probability that the top-ranked element changes grows significantly. Together, these results demonstrate that retrieval outcomes become highly sensitive to even small perturbations in the embedding space.

Importantly, this instability is not primarily driven by noise, but by the intrinsic geometric properties of high-dimensional representations. As shown previously, similarity scores concentrate and the separation between candidates collapses, so that small variations (arising from numerical precision, embedding variability, or minor input changes) are sufficient to alter the ordering of neighbors. In this regime, retrieval is no longer robust, but effectively operates near a decision boundary where multiple candidates are nearly indistinguishable.

\begin{figure}[htbp]
\centering
\includegraphics[width=0.45\textwidth]{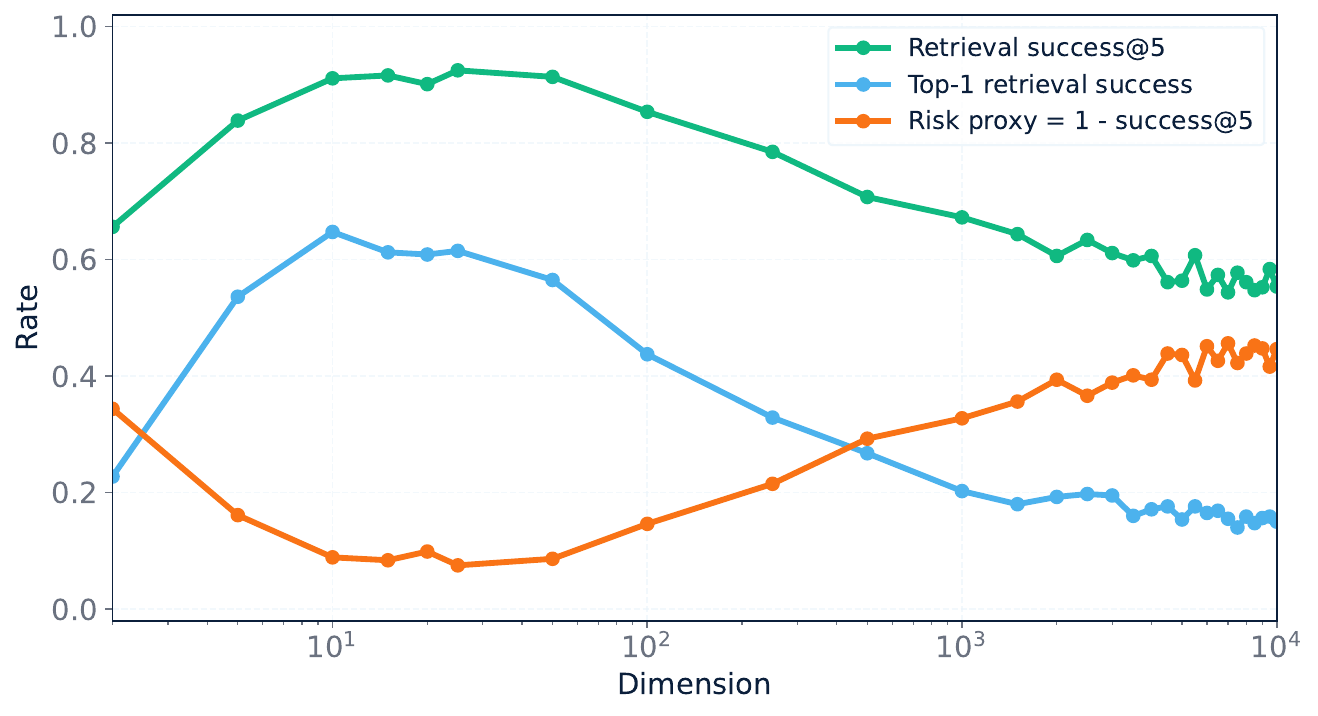}
\caption{(color online) Retrieval success and grounding degradation proxy as a function of dimension.}
\label{fig:rag}
\end{figure}

The practical consequences of this effect become evident when considering RAG systems. Fig.~\ref{fig:rag} presents the evolution of retrieval success rates alongside a proxy for hallucination risk, defined as the failure to retrieve relevant items within the top-$k$ results.

As the dimension increases, retrieval success degrades, while the risk proxy increases, indicating a growing likelihood that relevant information is not included in the retrieved context. This provides empirical evidence of a direct pathway from high-dimensional geometry to downstream system performance: when retrieval becomes unstable and unreliable, the generative model is forced to operate on incomplete or weakly relevant information.

From an operational standpoint, this establishes a geometric mechanism contributing to hallucination risk. The model itself may remain unchanged, but the quality of the retrieved context deteriorates due to concentration effects, leading to outputs that are coherent but insufficiently grounded in relevant data.

Taken together, these results highlight a critical insight: instability and performance degradation in RAG systems can emerge even in idealized settings, purely as a consequence of high-dimensional geometry. This underscores the need for robust retrieval strategies, additional validation layers, and diagnostic metrics capable of detecting and mitigating these effects in production environments.


\section{Mitigation and Diagnostic Strategies}
The results of this study indicate that degradation in similarity-based retrieval is not an implementation artifact, but a structural consequence of high-dimensional geometry. As such, mitigation does not consist in removing these effects, but in controlling their impact through appropriate design choices at multiple levels of the pipeline.

A first line of defense consists in improving the geometric conditioning of the embedding space. Centering and normalization are necessary but not sufficient, as they remove first-order bias while revealing concentration effects. More advanced techniques such as whitening, isotropy correction, or mean-direction removal can reduce anisotropy and mitigate the emergence of dominant directions. In particular, removing dominant embedding directions has been shown to improve isotropy and semantic discriminability in word representations~\cite{mu2018allbutthetop}. In practice, these transformations help restore a more balanced distribution of vectors and reduce spurious similarity induced by global biases.

A second approach is to move beyond single-metric similarity. Since both cosine similarity and Euclidean distance suffer from concentration, combining multiple signals can improve robustness. Hybrid retrieval strategies that integrate lexical matching, sparse representations, or structural constraints provide complementary information that is less sensitive to high-dimensional effects. In particular, re-ranking stages based on cross-encoders or task-specific scoring functions can reintroduce discriminative power after the initial retrieval step.

Another important lever lies in controlling the effective dimensionality of the representation. While increasing dimensionality improves expressivity, it also amplifies concentration phenomena. Dimensionality reduction techniques, such as principal component analysis or learned projections, can be used to identify and retain the most informative directions while discarding noise-dominated components. More generally, regularizing the embedding space to limit redundancy can improve the signal-to-noise ratio of similarity measures.

Given the observed instability of nearest-neighbor rankings, it is also essential to incorporate robustness checks into the retrieval process. Stability-based diagnostics, such as measuring the sensitivity of retrieved sets under small perturbations or tracking score gaps, provide direct indicators of retrieval reliability. These signals can be used to detect ambiguous queries, trigger fallback strategies, or adjust the number of retrieved candidates dynamically.

The presence of hubness further suggests the need for bias-aware retrieval mechanisms. Penalization schemes that downweight overly frequent neighbors, diversity constraints, or mutual proximity measures can reduce the dominance of hub vectors and improve the representativeness of retrieved results. Such techniques help ensure that retrieval reflects query-specific relevance rather than global structural bias.

At the system level, these considerations motivate the use of multi-stage retrieval architectures. Instead of relying on a single nearest-neighbor query, robust pipelines combine coarse retrieval, filtering, re-ranking, and validation steps. In RAG systems, this may include verifying the relevance of retrieved chunks, aggregating information from multiple candidates, or incorporating uncertainty-aware selection mechanisms before passing context to the generative model.

Finally, it is critical to monitor retrieval quality as a first-class component of system performance. Metrics such as contrast, variance, stability, and hubness should be tracked alongside traditional accuracy measures. In production environments, this enables early detection of degradation, supports debugging, and provides a basis for continuous improvement of the retrieval layer.

Overall, the mitigation of high-dimensional effects requires a shift from purely metric-based retrieval toward a more robust, multi-signal, and monitored approach. By explicitly accounting for geometric limitations, it is possible to design retrieval systems that remain reliable even in regimes where similarity measures alone are insufficient.


\section{Limitations}
The experiments in this paper are intentionally conducted under controlled synthetic conditions in order to isolate intrinsic geometric effects.

Real-world embedding systems additionally exhibit semantic structure, anisotropy, corpus-specific effects, model-dependent artifacts, tokenization and chunking effects. These factors may amplify or partially mitigate the observed behavior.

The objective of this work is therefore not to reproduce the full complexity of deployed systems, but to identify structural geometric mechanisms that can contribute to retrieval degradation. Importantly, the analyzed phenomena are not limited to embedding-based retrieval. Distance-based methods such as k-nearest neighbors, clustering algorithms, and kernel methods are also subject to concentration and instability effects in high-dimensional spaces.


\section{Conclusion}
This work provides a systematic and controlled analysis of how high-dimensional geometry affects similarity-based retrieval. The experiments consistently reveal that, as dimensionality increases, both distance and angular measures lose discriminative power. Pairwise similarities concentrate, relative contrast collapses, score gaps shrink, and nearest-neighbor structures become increasingly unstable and biased.

These effects are not artifacts of a specific model or dataset, but intrinsic properties of high-dimensional representation spaces. As a consequence, retrieval operates in a regime where candidate points become nearly indistinguishable, and ranking decisions rely on increasingly small and fragile differences. This leads to a structural form of instability, in which small perturbations are sufficient to alter retrieval outcomes.

The impact of this phenomenon becomes particularly significant in RAG systems, where retrieval serves as the interface between data and generation. The experimental results show that retrieval quality degrades with dimension, while a grounding degradation proxy increases. This establishes a concrete mechanism through which geometric effects can propagate to downstream system performance: when relevant information is not reliably retrieved, the generative model is forced to operate on incomplete or weakly grounded context.

From an operational perspective, this implies that part of the observed unreliability in AI systems may originate upstream, at the level of embedding geometry and retrieval, rather than solely within the generative model itself. Ensuring robust behavior therefore requires not only improvements in model architecture, but also a careful assessment of the geometric properties of the embedding space and their impact on similarity-based search.

More broadly, these results suggest that high-dimensional geometry should be treated as a structural component of AI system design. Monitoring concentration effects, retrieval stability, and structural biases such as hubness is essential for understanding and mitigating failure modes in embedding-based pipelines. While these effects cannot be eliminated, they can be measured and controlled, providing a foundation for more reliable and auditable AI systems.


\end{document}